# Self-Adaptive ERP: Embedding NLP into Petri-Net creation and Model Matching

Ahmed Maged
Business Informatics & Operations Management
The German University in Cairo
Cairo, Egypt
ahmed.abdel-hady@guc.edu.eg

Gamal Kassem
Business Informatics & Operations Management
The German University in Cairo
Cairo, Egypt
gamal.kassem@guc.edu.eg

**Abstract—** Enterprise Resource Planning (ERP) consultants play a vital role in customizing systems to meet specific business needs by processing large amounts of data and adapting functionalities. However, the process is resource-intensive, time-consuming, and requires continuous adjustments as business demands evolve. This research introduces a Self-Adaptive ERP Framework that automates customization using enterprise process models and system usage analysis. It leverages Artificial Intelligence (AI) & Natural Language Processing (NLP) for Petri nets to transform business processes into adaptable models, addressing both structural and functional matching. The framework, built using Design Science Research (DSR) and a Systematic Literature Review (SLR), reduces reliance on manual adjustments, improving ERP customization efficiency and accuracy while minimizing the need for consultants.

## I. INTRODUCTION

The implementation and customization of Enterprise Resource Planning (ERP) systems is a complex and resource-heavy process that significantly impacts organizations. ERP consultants are crucial for tailoring these systems to fit the specific needs of a business, involving an in-depth analysis of large datasets, mapping out business requirements, and adapting workflows to match ERP functionalities. This manual process requires extensive knowledge of both ERP technicalities and the organization's operations, making it a time-consuming and costly endeavor. As ERP systems are typically delivered in a sector-specific, but enterprise-neutral, fashion, consultants must customize these systems to align them with a company's unique workflows, which can take considerable effort [1].

Traditionally, ERP customization involves multiple phases, including the initial implementation and subsequent post-deployment modifications to accommodate changing business needs. As businesses evolve, the system must be continuously fine-tuned, leading to prolonged reliance on consultants. The complexity and rigid structure of these systems add to the cost and the duration of this process. Many enterprises spend substantial resources and time engaging ERP experts, which places a heavy burden on both time and finances [2].

The Self-Adaptive ERP Framework, proposed in this research, aims to alleviate these challenges by automating the customization process using AI. The framework leverages NLP and Petri nets to transform business process models into ERP system configurations, enabling real-time adjustments without the need for constant manual intervention. This automated approach allows businesses to implement and adjust ERP systems more efficiently, reducing their reliance on consultants and cutting down implementation times. One of the key innovations of this framework is its ability to bridge the gap between structural and functional model alignment in ERP systems.

Traditional ERP customization methods, such as those used by ERP consultants or older AI-based approaches like Markov Logic Networks or Case-Based Model Matching [3], often focus on structural alignment—ensuring that the system's architecture mirrors the organization's business processes. However, they fall short in addressing functional alignment, which takes into account the dynamic and contextual needs of the business. By incorporating both structural and functional matching, the Self-Adaptive ERP Framework ensures a more holistic alignment between the system and the business processes.

The solution follows the Design Science Research methodology combined with a Systematic Literature Review to build and validate the framework. It introduces a novel AI-driven approach that continuously adapts the ERP system based on real-time data, adjusting to the evolving needs of an organization. This dynamic capability is key to reducing not just the implementation time, but also the ongoing maintenance costs associated with ERP systems.

The framework's use of NLP allows it to interpret and transform business process models into actionable ERP configurations, further enhancing its adaptability. Petri nets, which provide a mathematical representation of business processes, are used to model the workflows and ensure that the system remains aligned with operational changes.

A core benefit of this framework is its ability to reduce labor costs and the time-intensive work traditionally carried out by consultants. Rather than requiring weeks or months of back-and-forth adjustments between the consultant and the business [1].

This research on ERP system customization is essential due to the significant challenges faced during implementation, particularly for small and medium-sized enterprises (SMEs). These challenges include resource constraints, limited IT expertise, and high customization demands, which often result in project failures. Studies indicate that between 70% and 85% of ERP implementations encounter issues such as cost overruns, timeline delays, or misalignment with business processes. Additionally, only about 33% of ERP projects are considered fully successful, demonstrating the complexity and risks involved in such undertakings. [4][5]

Where the Self-Adaptive ERP system automatically learns from the organization's existing processes and adjusts itself to meet those needs. This significantly accelerates the ERP implementation process while ensuring that the system is continually optimized as the business evolves.

This research also tackles a broader gap in the field: the lack of AI-driven customization solutions in ERP implementations. While ERP systems have become essential in managing core business operations, they remain difficult to customize without extensive manual intervention. The Self-Adaptive ERP Framework addresses this gap by introducing a more flexible, responsive, and intelligent approach to ERP customization. It uses AI not only to align the system's structure with the business processes but also to understand the context and function of these processes, ensuring that the system adapts in real time as organizational needs change, offering a more efficient, cost-effective, and dynamic approach to managing enterprise systems.

## II. Related Work

The gap is shown more specifically in the work of [6] where the author explores how SAP integrates AI technologies to enhance its functionalities. Key areas of focus include the use of machine learning for predictive maintenance and demand forecasting, enabling the system to continuously improve decision-making by recognizing patterns in historical data. NLP allows SAP to understand user queries and perform sentiment analysis, fostering intuitive interactions. AI-powered chatbots provide real-time assistance, simplifying processes, while computer vision automates tasks like invoice processing. The study highlights how these AI-driven innovations contribute to a more efficient, automated, and adaptive ERP system.

Moreover, the paper of [7], delves into the examination of AI biases within the context of ERP software customization. It underscores the significance of recognizing how AI algorithms used to optimize ERP systems may introduce unintended biases that affect decision-making processes. The study explores the application of models like the Prioritized Requirements Customization Estimation (PRCE) and K-nearest-neighbors ERP (KNN) algorithms, both of which forecast the level of customization needed for ERP systems based on client requirements. By focusing on these algorithms, the authors reveal how biases in customization models can lead to inefficiencies, affecting operational performance and employee experiences. The paper also discusses proactive strategies to mitigate these biases, emphasizing the need for fairness, transparency, and accountability in AI-driven ERP customization processes.

To test out the AI advancements in ERP, [8] developed a vocal NLP interface to enhance ERP systems by converting spoken commands into SQL queries, bridging user-friendly interaction with database precision. For instance, a verbal request like "Show all pending orders for October" is translated into an SQL statement to retrieve relevant data, facilitating easier access for non-technical users and reducing reliance on IT teams. This system was applied in a manufacturing context, effectively extracting operational metrics and key performance indicators. While the approach showed promise, challenges such as managing ambiguities in user queries, supporting complex query structures, and ensuring compatibility with diverse database schemas persist. Although, further research is needed to improve the system's scalability, generalizability across industries, and its ability to handle errors, making it more robust for various organizational scenarios.

Where [9] examined how generative AI enhances data migration and conversion processes during ERP SaaS deployments. The approach involves preprocessing enterprise data, aligning it with ERP schemas, and using generative models to resolve inconsistencies, automate schema mapping, and validate data integrity. These AI-driven methods improve scalability, accuracy, and efficiency, minimizing manual effort and enabling real-time adjustments in cloud environments. While generative AI offers clear advantages over traditional methods, such as handling large and diverse datasets with fewer errors, it also faces challenges. Limitations include dependency on high-quality training data and difficulties in tailoring models to specific organizational needs. The study underscores the transformative potential of generative AI for data conversion while advocating further refinement of these techniques for broader, more adaptable ERP applications.

However, [1] presented a forward-thinking concept for enhancing ERP systems through the integration of data mining and AI methods. The authors proposed a self-adaptive system that autonomously adjusts its customization in response to real-time insights, contrasting traditional static methods. They focus on two phases: pre-implementation and post-implementation. In the pre-implementation phase, ERP consultants use reference models as patterns to match ERP systems with business processes.

A Self-Adaptive Customizing Model (SACMS) leverages text mining to translate business terms into ERP terminology, enabling the alignment of enterprise workflows with ERP functionalities. The post-implementation phase uses trace files to monitor user interactions with the system, offering insights into workflow execution. These traces are then analyzed to continuously improve the ERP system's fit with business needs, fostering an agile, data-driven customization process.

The literature surrounding self-adaptive ERP systems and AI enforced ERP systems has highlighted significant advancements in extending ERP functionalities. However, two critical challenges have emerged. First, the lack of AI involved during the ERP customization itself where ERP consultants struggle to map the business terminologies and business process to the ERP terminologies and process.

One of the main concerns that these processes face is the terminology problem which is how the ERP and business domains interact. The ERP domain and the business domain is mapped to its equivalent in the formal model's terminology after many iterations and meetings just to find an equivalent corresponding synonym for each word in the business process and the ERP reference models [10]. Secondly, even after the terminologies are mapped, the model matching between business process and ERP reference models is a long and cost intensive process as the constructed business process is model matched with ERP reference models in order to ensure that the tailored ERP solutions adequately reflect the unique workflows and operational needs of an organization while also adhering to best practices established in the reference models.

However, traditional model matching approaches predominantly emphasize structural alignment between business and ERP models, often neglecting the semantic context and linguistic nuances embedded in the business processes. This focus on structural matching alone overlooks critical differences in terminology, phrasing, and domain-specific language, which can lead to suboptimal alignment. The lack of attention to these contextual elements may result in a misinterpretation of business requirements, ultimately affecting the customization and implementation of ERP systems. Addressing both the structural and semantic dimensions of model matching is essential for a more holistic and accurate alignment between business processes and ERP reference models, ensuring that the system captures not only the workflow but also the intent and specificities of the business environment.

### III. RESEARCH METHODOLOGY

The paper's primary objective was to develop a Self-Adaptive ERP system. This goal was addressed through a conceptual model created using the Design Science Research methodology along with a Systematic Literature Review. The research question, "How can NLP and semantic model matching enhance the alignment of Petri-Net models with ERP reference models for automated ERP customization?" guided the study. The resulting framework integrates NLP and Petri nets to automate ERP customization, aligning business processes with ERP reference models and improving both structural and functional model matching. This framework aims to automate and streamline ERP customization, reducing the need for manual adjustments while ensuring that the system remains adaptive to evolving business processes.

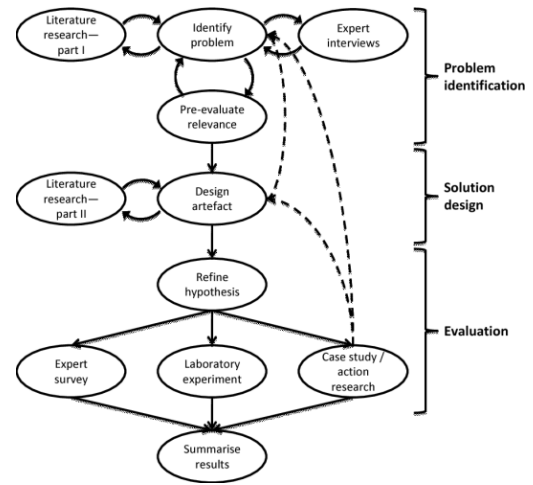

Fig. 1. Offerman's Design Research Process [11]

Figure 1 illustrates Offerman's Design Research methodology with three main steps of Problem Identification, Solution Design and Evaluation.

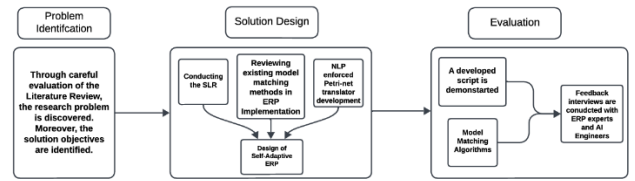

Fig. 2. The Solution Approach of Self-Adaptive ERP

Figure 2 illustrates the steps for the solution approach of the developed framework.

#### A. Problem Identification

The proposed self-adaptive ERP aims to tailor systems to specific business needs using unstructured text, which describes business objects and relations. By leveraging NLP, the framework translates business terminology and processes into ERP models, enabling efficient and context-aware ERP customization through Petri-Net creation and model matching.

#### B. Solution Design

To address the research problem, the development of a Self-Adaptive ERP follows three main design steps. The first step involves conducting a Systematic Literature Review to identify various approaches used in model matching. The second step is reviewing existing model matching frameworks to gather insights. The third step focuses on developing an NLP-powered Petri-Net translator, which facilitates automated ERP customization.

#### C. Evaluation

The evaluation of the Self-Adaptive ERP Framework was conducted in two stages. First, the business process was translated into a Petri-Net using NLP. In the second stage, the generated Petri-Net was matched with an ERP reference model through mathematical operations, also using NLP.

Finally, expert interviews were conducted with two representatives from academia and two from the industry. These experts assessed the framework's effectiveness after reviewing the demonstration and scenarios, which involved translating business processes and matching them with ERP models.

IV. RESULTS

This research tackles the need for AI involvement in ERP systems specifically the customization phase, where the terminology mapping and model matching are crucial in order to fulfil a best-fit ERP systems satisfying the business requirements where results of this research is demonstrated as follows:

*A. Problem Definition*

The review of existing research on NLP and machine learning techniques for ERP system customization reveals significant gaps, particularly the absence of a comprehensive framework that unifies enterprise-specific and ERP terminologies. This study aims to develop a novel, adaptive model matching approach that leverages advanced techniques to enhance the scalability and effectiveness of ERP customization solutions.

*B. Solution Design*

The research solution's design was formulated of two main steps:

1. Conducting the Systematic Literature Review: A SLR was initialized according to [12] methods in order to summarize the literature review regarding ERP implementation and customization where it resulted in 25 related articles which had the question of "What is the currently used model matching methods carried out in ERP customization?"

2. Reviewing of existing model matching: The review was mainly conducted in order to obtain and use the most advanced techniques available found in the literature review. In total 6 methods and techniques were selected to be reviewed: 1 method related to the structure or arrangement of elements within process models, 1 technique matching solely relying on labels, 1 method approach that allows for a sophisticated, probabilistic treatment of process model matching, and 2 methods that incorporate the use of machine learning and NLP into model matching.

This SLR has provided the research of the methods needed to model match the business process to the extracted ERP reference models to reach a coherent ERP implementation which satisfies the business needs.

The proposed framework integrates NLP and Petri-Net technologies to enable self-adaptive ERP customization through three key technical contributions. First, it uses pre-trained NLP models like BERT and Word2Vec to semantically analyse business process descriptions, capturing contextual relationships and representing process elements as high-dimensional embeddings. Second, it translates these embeddings into Petri-Net elements (places, transitions) while preserving structural dependencies, such as loops and decision points, ensuring workflow accuracy. Finally, the framework incorporates iterative alignment mechanisms, leveraging semantic similarity measures (e.g., LIN and WordNet path lengths) and structural mapping to resolve discrepancies between business workflows and ERP reference models, enabling real-time adaptation with minimal manual intervention.

*C. Evaluation*

The Evaluation phase however was formulated of two main steps:

1. NLP Petri-Net translator: A script was developed to handle business processes and formulate a workflow network Petri-Net that can be used later to match and map with an ERP reference model.

2. Model Matching: A cosine Similarity was conducted to map out differences between the developed Petri-Net and the ERP reference models

Various business process examples were tested to ensure the functionality of the NLP Petri-Net translator.

1. The source code was designed using two distinct algorithms to handle Petri-Nets. the algorithm utilized SpaCy as the NLP library along with word embedding techniques to convert business process descriptions into vectors. The Matplotlib library was employed to visualize the resulting Petri-Net workflows, which include tokens, functions, and actions.

    The functions were specifically designed to manage both branching and non-branching cases. Figure 3 below detail a snippet of the code structure, with clear annotations explaining each component. The algorithm's outputs are depicted in Figure 4 which show the results of inputting "The customer places an order, the system checks inventory. If the stock is available, the order is confirmed and packed. If the stock is not available, a purchase order is created, goods are received, and then the order is shipped to the customer." as a business process.

```
import spacy
import networkx as nx
import matplotlib.pyplot as plt
nlp = spacy.load("en_core_web_sm")
def extract_actions(text):
    doc = nlp(text)
    actions = []
    for token in doc:
        if token.pos_ == "VERB":
            present_verb = token.lemma_.capitalize()
```

```
    actions.append(present_verb)
  return actions
def extract_sentences(text):
  doc = nlp(text)
  sentences = [sent.text for sent in doc.sents]
  return sentences
def extract_conditions(sentences):
  conditions = []
  for sent in sentences:
    if "if" in sent.lower():
      conditions.append(sent.strip())
  return conditions
def create_petri_net_with_conditions(actions, sentences, conditions)
```

Fig. 3. NLP-Enforced Petri-Net Translator Code Snippet

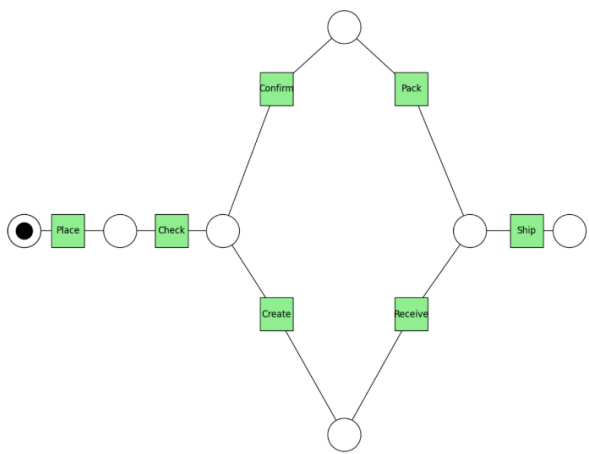

Fig. 4. NLP-Enforced Petri-Net Translator Result

2. Model Matching: Afterwards, the next step in the Self-Adaptive ERP is to match the developed Petri-Net model with and ERP reference model and compare both models to find the best-fit model for implementation using the SLR results. This was developed through cosine similarity algorithms, embedding difference and structure similarity using GloVe word embedding model to vectorize objects. The model matching was conducted through comparing a provided order fulfillment business model and 3 extracted ERP reference models.

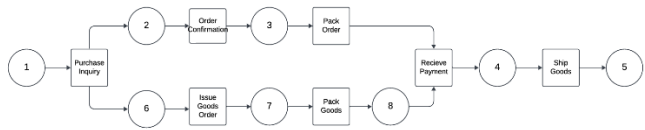

Fig. 1. Business Model R1

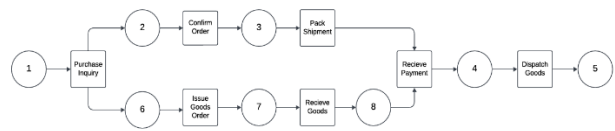

Fig. 6. ERP Reference Model R2

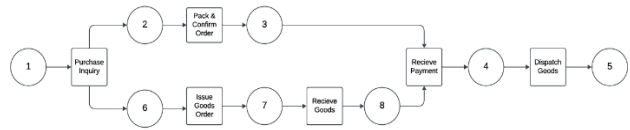

Fig. 7. ERP Reference Model R3

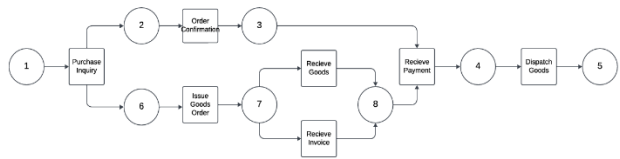

Fig. 8. ERP Reference Model R4

These three ERP reference models (R1, R2, R3) were matched with the developed Petri-Net business model (fig. 4) using Petri nets and mathematical similarity metrics. Tasks were vectorized using GloVe embeddings to compute embedding similarity, while cosine similarity and structural similarity analyzed alignment between models.

For instance, R1 embeddings showed a strong match (cosine similarity: 0.97) due to minimal deviations in terminology and structure. While R3 and R4 exhibited more variations, leading to lower similarities (cosine: 0.857 and 0.928). Structural similarity reflects workflow alignment, slightly reduced by task modifications or additions. This illustrates Petri nets' utility in precise ERP-process matching. To elabarote, the calculations were done as follows:

For R2 and the business model (fig. 4):

- Vectorizing:
  - $V_1 = [0.8, 0.9, 0.7, 0.85, 0.9]$,
  - $W_1 = [0.81, 0.89, 0.71, 0.86, 0.91]$

- Cosine Simialrity

$$\frac{V_1 \cdot W_1}{\|V_1\| \, \|W_1\|} \frac{3.496}{1.897 \times 1.898} = 0.97$$

This quantitative evaluation facilitates informed ERP selection by assessing semantic and structural congruence.

The final results show that ERP Reference Model R2 has the highest similarity to the business model, with an embedding similarity of 0.970 and structure similarity of 1.0. ERP Model R4 follows closely, with an embedding similarity of 0.928 and structure similarity of 0.9, while ERP Model R3 has the lowest similarity due to additional steps, achieving only 0.857 in embedding and 0.85 in structure. Models R2 and R4 demonstrate around 95% similarity to the corporate model, whereas R3 falls to 85% due to parallel functions.

## V. CONCLUSION

This study presents a transformative approach to ERP system customization through a self-adaptive framework integrating NLP technologies and Petri-Nets. The framework demonstrated its ability to streamline ERP implementation by enabling dynamic, real-time customization, significantly reducing manual effort. Key contributions include the use of word embeddings and semantic matching for accurate model alignment, and a novel NLP-driven Petri-Net translator to automate workflow processes. The findings validate the framework's potential to improve organizational agility and adaptability by facilitating efficient ERP customization.

This framework was designed to be integrated with ERP systems, ensuring that business processes can be automatically translated, transformed, and matched. For instance, ERP system experts and consultants can use this framework to compare business process data from the ERP reference models, including process comparison and customization analytics. Moreover, the framework serves as an input to process analysis tools, enabling various types of process analysis, such as model matching and workflow optimization. As, this approach provides precise model matching, aiding in ERP customization by identifying both suitable reference models and key adjustment areas during implementation.

However, challenges such as scalability in large-scale ERP systems and a lack of real-world testing highlight areas for improvement. Future research should focus on enhancing scalability through advanced machine learning techniques and validating the framework across diverse industries like manufacturing, healthcare, and finance. Refining the NLP Petri-Net translator to address sector-specific nuances would also boost its accuracy and versatility.

Moreover, incorporating real-time monitoring of user behaviors could enable the framework to dynamically adjust workflows, ensuring continuous improvement and responsiveness to evolving business needs. These advancements would further solidify the framework's role in revolutionizing ERP systems, making them more adaptive and aligned with modern organizational demands.

Additionally, refining the NLP Petri-Net translator to handle sector-specific nuances would improve its accuracy and versatility. But, to achieve full self-adaptability, continuous monitoring of user behaviors during daily operations would allow the system to dynamically adjust workflows, eliminate inefficiencies, and respond to changing business needs in real time.

For example, if the system observes that certain functions are frequently bypassed or not fully utilized, it can recommend or even implement streamlined configurations, removing unnecessary steps or providing tailored suggestions to users. Similarly, if user behaviours indicate a change in workflow priorities (e.g., a shift in order processing due to seasonal demand), the system could adapt by reallocating resources or modifying workflows automatically.

This continuous monitoring and adaptability can significantly enhance the ERP customization phase. Instead of relying solely on initial setups or periodic manual adjustments, the system can fine-tune itself in real time, leading to more responsive and agile operations. This is particularly important in industries where processes are constantly evolving, and businesses need their ERP systems to adapt swiftly without the need for constant intervention from consultants or IT professionals.

Ultimately, a self-adaptive ERP system becomes a living, evolving entity that responds not only to predefined rules and configurations but also to the actual usage patterns and changing requirements of the business, ensuring more efficient, cost-effective, and personalized process management. This would make the ERP system more responsive, agile, and efficient, evolving alongside the organization without requiring constant manual intervention.